\documentclass[aastex]{emulateapj}
\usepackage{apjfonts} 
\usepackage{psfig}
\usepackage{graphicx}
\shorttitle{ULTRALUMINOUS X-RAY SOURCES}
\shortauthors{KALOGERA, HENNINGER, IVANOVA, \& KING}

\begin{document}

\title{An Observational Diagnostic for Ultraluminous X-Ray Sources}

\author{V.~Kalogera \altaffilmark{1}, M.~Henninger\altaffilmark{1},
N.~Ivanova\altaffilmark{1}, and A.~R.~King\altaffilmark{2}}

%\altaffiltext{1}{Northwestern University, Dept. of Physics \& Astronomy, 2145 Sheridan Rd., Evanston, IL 60208; vicky@northwestern.edu, m-henninger@alumni.northwestern.edu, nata@vardar.phys.northwestern.edu}
%\altaffiltext{2}{University of Leicester, Leicester, LE1 7RH UK; ark@star.le.ac.uk}

\affil{ $^{1}$ Northwestern University, Dept. of Physics \& Astronomy, 2145 Sheridan Rd., Evanston, IL 60208\\ $^{2}$ University of Leicester, Leicester, LE1 7RH UK.\\ vicky@northwestern.edu, m-henninger@alumni.northwestern.edu, nata@vardar.phys.northwestern.edu, ark@star.le.ac.uk}

 \begin{abstract} We consider observational tests for the nature of
Ultraluminous X-ray sources (ULXs). These must distinguish between
thermal--timescale mass transfer on to stellar--mass black holes
leading to anisotropic X-ray emission, and accretion on to
intermediate--mass black holes. We suggest that long--term transient
behavior via the thermal--viscous disk instability could discriminate
between these two possibilities for ULXs in regions of young stellar
populations. Thermal--timescale mass transfer generally produces
stable disks and persistent X--ray emission. In contrast, mass
transfer from massive stars to black holes produces unstable disks and
thus transient behavior, provided that the black hole mass exceeds
some minimum value $M_{\rm BH,min}$. This minimum mass depends
primarily on the donor mass and evolutionary state.  We show that
$M_{\rm BH,min} \gtrsim 50$\,M$_\odot$ for a large fraction ($\gtrsim
90$\%) of the mass--transfer lifetime for the most likely donors in young
clusters. Thus if long--term monitoring reveals a large transient
fraction among ULXs in a young stellar population, these systems would
be good candidates for intermediate--mass black holes in a statistical
sense; information about the donor star is needed to make this
identification secure in any individual case. A transient ULX
population would imply a much larger population of quiescent systems
of the same type.

 \end{abstract}


\section{INTRODUCTION}
\label{sec:intro}

In the past few years high-angular--resolution observations with {\em
Chandra} have revolutionized the study of X--ray binaries in nearby
galaxies and have revealed whole populations of sources in a variety
of galaxy types~\citep[for a recent review see][]{FW2003}. The
detected X--ray fluxes have been combined with distance estimates to
the host galaxies to infer the {\em apparent} X--ray luminosities of
the sources, {\em assuming isotropic} emission. The inferred X--ray
luminosities reveal a distinct class of sources: non--nuclear point
sources with apparent X--ray luminosities above the Eddington
limit for a $\sim 10$\,M$_\odot$ black hole ($\gtrsim
2\times 10^{39}$\,erg\,s$^{-1}$), often referred to as {\em ultraluminous X-ray sources} (ULXs). The existence of such sources was first noted in EINSTEIN observations~\citep[e.g.,][]{Fab88}. Short-term variability detected in a number of them~\citep[see e.g.,][]{Mats2001,FZ03} excludes the
possibility of source confusion and strongly points towards
accretion as the origin of the X-rays. At present the majority of ULXs
have been found mainly in young stellar populations and regions of
recent star formation, although a few have been identified in
elliptical galaxies~\citep[e.g.,][]{SIB,CP} with luminosities close to
the lower end of the ULX range. 

If the apparent X-ray luminosities are indeed the true luminosities of
the sources, their high values have very important implications for
their accreting compact objects. For sources with X-ray luminosities comparable or in excess of $10^{40}$\,erg\,s$^{-1}$, the Eddington limit gives a lower limit on the mass intermediate between stellar ($\lesssim 50$\,M$_\odot$) and supermassive ($\gtrsim 10^6$\,M$_\odot$) black
holes (BH). ULXs may thus suggest the existence of a new class of compact objects: {\em intermediate-mass black holes} \citep[IMBH;][]{CM99}.

On the other hand it is still possible that the accreting compact
objects in ULXs are of stellar mass \citep[$\lesssim 20$\,M$_\odot$;
see][]{BKB}. The high apparent X-ray luminosities can be explained in
two different ways: (i) either the Eddington limit (rigorously derived
for spherical accretion) is not relevant and in fact can be exceeded
\citep[see][]{RB} or (ii) the apparent X-ray luminosities overestimate
the true source luminosities because the emission is {\em
anisotropic}~\citep{King}. Although the theoretical basis for
imposing the Eddington limit is somewhat unclear, there is strong
support for it partly from observations of X-ray bursts from
accreting neutron stars~\citep[e.g.,][]{erik,LvPT} and from the current understanding of the evolutionary history of wide binary pulsars~\citep{WK} and Cygnus X-2 \citep[where the compact object does not seem to have gained any significant amount of mass;][]{KR99,Kolb,PR}.

Anisotropic emission is probably associated with X-ray luminosities
comparable to the Eddington limit. Binary systems can reach such high
luminosities in two different situations~\citep{K2002}:
(i) thermal-timescale mass transfer typically occurring when
the donor is more massive than the accretor~\citep{King}. 
Cygnus X-2~\citep{KR99} and SS433~\citep{KTB} may be
examples of this phase;
(ii) X-ray transient outbursts, where the 
thermal disk instability governs the accretion behavior. The first
possibility obviously requires donors more massive than black holes
($\gtrsim 3-5$\,M$_\odot$), and hence relatively young stellar
environments, whereas the second must apply to ULXs in
old elliptical galaxies \citep{PB}.

Although population studies suggest that a large fraction of ULXs must
be stellar--mass X--ray binaries~\citep{GGS}, some may
contain IMBH. For the ULX in M82, the very
high peak luminosity~\citep{Mats2001}, the quasi-periodic oscillations~\citep{SM}, and the detection of an isotropic nebula around it may
point away from the anisotropic-emission possibility~\citep[although
see][]{KingPounds}. On the other hand, no ULXs are found inside dense
clusters, where IMBH are expected not only to form~\citep{PZMcM,MH}
but also remain, as they are much heavier than the average stellar
mass in clusters~\citep[fast cluster disruption could help, but this
issue is beyond the scope of this paper;][]{GR}.  Thus at present the
physical origin of some of these sources is not clear, and there may
be ULXs of both stellar and intermediate mass.

In this {\em Letter} we suggest that long-term transient behavior (not just flux variability by factors of a few to several) 
due to the thermal-viscous disk instability~\citep{KKB,KR98} may distinguish the two possibilities for ULXs in regions of young stellar populations ($\lesssim 10^{8}$\,yr). We show that one can define a minimum BH mass $M_{\rm BH,min}$ for
disk instability and thus transient behavior (\S\,2). This minimum
mass depends primarily on the mass and evolutionary stage of the donor
star. We show that, for donor masses $\gtrsim 5$\,M$_{\odot}$ (expected to be the most likely donors in young stellar
environments), $M_{\rm BH,min} \gtrsim 50$\,M$_\odot$ for a large
fraction of the mass-transfer phase ($\gtrsim 90$\%).  (\S\,3). By contrast, thermal timescale mass transfer is expected to be persistent~\citep{King}.  Thus if long--term monitoring reveals a significant transient fraction among ULXs in a young stellar population, these systems would be good candidates for
IMBH in a statistical sense; information about the donor star is needed to make this identification secure in any individual case. In \S\,4 we discuss the observational significance of this diagnostic and the connection to IMBH formation scenarios.

\section{MINIMUM BLACK HOLE MASS FOR TRANSIENT BEHAVIOR} 

The thermal--viscous disk instability provides a currently accepted
explanation for transient behavior in X--ray binaries. The instability
causes the disk to undergo a limit cycle in which the central
accretion rate passes through short high states (outbursts) and long
low states (quiescence). This picture was originally developed to
explain dwarf novae but can be extended to soft X--ray transients by
including the effects of disk irradiation~(van Paradijs 1996; King, Kolb \& Burderi 1996; Dubus et al.\ 1999; for reviews see Lasota 2001;
Frank et al.\ 2002). The condition for transient behavior is that the
disk surface temperature at its outer edge should lie below the
hydrogen ionization temperature. This in turn requires the mean mass
transfer rate to lie below a critical value $\dot{M}_{\rm crit}$~\citep{KKB}. This depends primarily on the binary component masses $M_{\rm BH}$, $M_2$ and orbital period $P$, and to a lesser degree on the detailed vertical disk structure. The latter is of course uncertain; in this paper we use the form given in
~\citep[eq.\ 32 in][]{Dubus}:
 \begin{eqnarray} 
\dot{M}_{\rm crit} & \simeq & 6.6\times 10^{-5}\,M_{\odot}{\rm
 yr}^{-1}\left(\frac{M_{\rm BH}}{100M_\odot}\right)^{0.5} \nonumber  \\
& & \times \left(\frac{M_2}{10M_\odot}\right)^{-0.2}\left(\frac{P}{1\,{\rm
 yr}}\right)^{1.4}.
 \end{eqnarray} 
 Although the precise conditions assumed by~\cite{Dubus} (in
particular the central mass and vertical structure) probably cannot be
extrapolated to all of the cases we shall consider, this equation
gives an adequate idea of when transient behavior is likely.  Using a
somewhat simpler expression~\cite{KKB} first showed that the condition
$\dot{M} < \dot{M}_{\rm crit}$ translates into a {\em minimum BH mass}
$M_{\rm BH,min}$ required for the development of transient
behavior. Similarly, equation (1) can be used to derive this minimum:
 \begin{eqnarray}  
 M_{\rm BH} & \gtrsim & 
 230M_{\odot}\,\left(\frac{\dot{M}}{10^{-4}M_{\odot}{\rm
 yr}^{-1}}\right)^{2}\nonumber \\ & & \times \left(\frac{M_2}{10M_\odot}\right)^{0.4}\left(\frac{P}{1\,{\rm
 yr}}\right)^{-2.8}.
\end{eqnarray} 
 
Our aim is to examine whether transient behavior favors a distinct BH mass range. We use mass transfer sequences calculated for a set of
initial binary configurations (of varying orbital periods, black hole and donor masses) and we derive $\dot{M}_{\rm crit}$ for a
given donor mass $M_2$ and radius $R_2$ (i.e., evolutionary state).
We then use the dependence of $\dot{M}_{\rm crit}$ on $M_{\rm BH}$ and disk
radius given in~\citep[eq.\ 30 in][]{Dubus} and solve numerically for
$M_{\rm BH,min}$ by setting $\dot{M}_{\rm crit}$ equal to the mass
transfer rate found from our mass transfer sequences, for given $(M_2,
R_2)$. If the BH mass used in the mass transfer calculations exceeds
this minimum the system will be transient. (Note that for a given
sequence, $\dot{M}$ depends most sensitively on the donor mass and
evolutionary stage at the onset of mass transfer and not so much on
the accretor mass.) We examine the range of values for $M_{\rm BH,min}$ and whether transient behavior can be associated with certain types of BH accretors (\S\,3.2).

\section{MASS TRANSFER SEQUENCES} 

\subsection{Stellar Evolution Code}

We calculate stellar models and mass-transfer sequences with an
updated stellar evolution code described in detail in
\citep{Pod,Iva}. The current version has been modified to minimize
numerical noise in the mass transfer calculations and ensure that the
stellar and Roche lobe radii track one another during mass
transfer. We use mixing length and overshooting parameters of 2 and
0.25 pressure scale heights respectively.  Since we are dealing with
massive stars, we account for mass loss due to stellar winds
\citep[rates adopted from][]{Hurley}. In calculating orbital changes
we take account of both mass transfer and wind mass loss with the
specific angular momentum of the mass-losing donor. We assume that any
mass transfer above the Eddington rate is lost from the binary with
the specific angular momentum of the accretor.

We model mass transfer self--consistently following
the donor response to the appropriate rate of mass loss. The
mass-transfer rate $\dot M$ is calculated in an {\em implicit} manner,
so that the donor radius $R$ remains equal to the Roche lobe radius
$R_{\rm L}$ \citep[using Eggleton's approximation;][]{Egg}. We
consider the radius-mass exponents of the Roche lobe $\zeta_{\rm L} =
d \ln R_{\rm L}/d \ln M$ and of the star itself $\zeta= d \ln R/d \ln
M$ in our solution method. The response of the Roche lobe to the mass
transfer is solely a function of the mass ratio, whereas the response
of the stellar radius depends on the mass transfer rate. For a given
model, we tabulate values of $\zeta$ for a range of $\dot{M}$
values. We then identify the value of $\dot{M}$ for which the Roche
lobe radius is equal to the stellar radius (predicted from the value
of $\zeta$). In some cases, the solution for $\dot{M}$ is not unique;
we then choose the lowest value to avoid large excursions in the
rate. As the donor evolves, the stellar-radius response changes, so we recalculate the table of $\zeta(\dot M)$, if the
predicted stellar radius differs from the calculated one by $\delta
\ln R = 10^{-4}$.

\subsection{Calculations and Results} 

To investigate the systematics of mass transfer and disk (in)stability we consider a large set of mass-transfer sequences driven by Roche-lobe overflow: binaries with BH masses in the range $10-1000$\,M$_\odot$, donor masses in the range $1-25$\,M$_{\odot}$. 
For each donor mass, we evolve single-star models to
different evolutionary stages that cover most of the stellar
lifetime: from the Zero-Age to the End of the Main Sequence (ZAMS and
EMS), the Hertzsprung gap (HG), and through core helium burning. We
consider the possibility of multiple mass-transfer episodes in the evolutionary history of each binary and we evolve each of our models up to carbon ignition.

\begin{figure}[!h]
\begin{center}
\psfig{figure=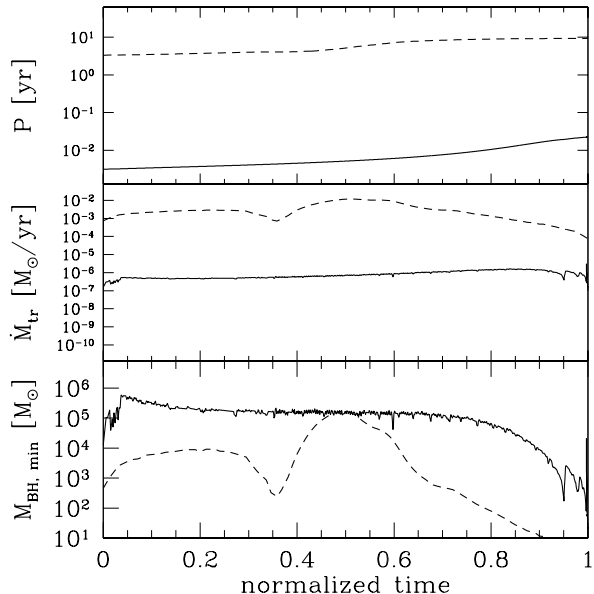,width=3.2in}
\caption{Two examples of mass-transfer sequences for a 1000\,M$_\odot$
BH with a 20\,M$_\odot$ donor. Mass transfer starts on the
Zero-Age Main Sequence (solid) and the Base of the Giant Branch (dotted). The
orbital period (top), mass transfer rate (middle), and derived minimum
BH mass for transient behavior (bottom) are shown as functions of
time normalized to the total duration of each mass transfer episode.}
\label{fig:mt}
\end{center}
\end{figure}

%\vspace{-0.6cm} 

Every evolutionary sequence of course starts and ends with
transient behavior as the mass transfer rate rises from and returns to
zero. For episodes of otherwise persistent mass transfer these
transient windows form a very small fraction of the total mass
transfer lifetime and have very low discovery probability. In terms of the solutions for $M_{\rm BH,min}$, this means that at the start and end of every mass-transfer episode the minimum BH mass for transient behavior is very low and certainly enters the stellar-mass range. We
eliminate these insignificant (due to their low discovery probability) transient epochs by excluding the first and last 5\% of the mass-transfer lifetime.

The behavior of two example mass-transfer sequences is shown in Figure
1. These have been calculated for $M_{\rm BH} =$ 1000\,M$_\odot$ and donors of $20$\,M$_\odot$ at two evolutionary stages: unevolved (ZAMS) and at the base of the Giant Branch. Our results for the binary orbital period, mass transfer rate, and minimum BH mass for transient behavior are shown as a function of time normalized to the total duration of each mass-transfer episode ($\simeq 10^7$\,yr and $\simeq 2\times 10^3$\,yr, respectively). Evidently, for most of these episodes (90\%-100\% of their duration), transient behavior requires BHs in the intermediate--mass range $M_{\rm BH,min} >$ 50\,M$_\odot$ and {\em not} the stellar--mass range. In reality, for stellar-mass binaries the orbital separation is small enough that the radiation field of the O,B donor is able to keep the disk ionized, and therefore stable. This effect is negligible for the much wider separations of IMBH binaries.

For sequences with donors down to 10\,M$_\odot$ and 7\,M$_\odot$ results are very similar to the 20\,M$_\odot$ sequences, with $M_{\rm BH,min} >$ 50\,M$_\odot$ for more than 90\% of their duration.  We note that ULX luminosities would be reached in outbursts and would then probably reflect the Eddington luminosity rather than the mass-transfer rates shown in Figure 1.

For binaries with more evolved donors (during most of the short phase of core-helium burning when orbital periods are $\simeq 10$\,yr), $M_{\rm BH,min}$ values can be $\lesssim 10$\,M$_\odot$. In these cases accretion disks are so large that their edges are cool and allow the disk instability to develop. However, such binaries are expected to be uncommon in young clusters for two reasons. First, they are too wide to survive stellar interactions; typical interaction timescales are $10^5$\,yr for stellar densities of $10^5$\,pc$^{-3}$~\citep[typical of young stellar clusters in star-forming regions; see \S\,8.4 in][]{BT}. Second, our numerical calculations show that their mass-transfer episodes are much shorter ($\lesssim 10^4$\,yr) than for MS donors ($10^5-10^7$\,yr). Therefore for the same formation probability, it is more difficult to detect such short-lived X-ray phases.  

Sequences with 5\,M$_\odot$ donors at different evolutionary stages show a qualitative change in behavior. They straddle along the dividing line between $M_{\rm BH,min}$ values in the intermediate-mass and the stellar-mass range because of a close balance between two effects on  $M_{\rm BH,min}$ (eq.\ [2]):  the orbital period and the mass trasfer rate at Roche-lobe overflow. As a result a 5\,M$_\odot$ donor at the ZAMS {\em and} beyond the MS gives $M_{\rm BH,min}$ values in the stellar-mass, whereas the same donor filling its Roche-lobe close to the end of the MS leads to $M_{\rm BH,min}$ values in the intermediate-mass range. Sequences with less massive donors always have $M_{\rm BH,min}$ values in the stellar-mass range, i.e., $<$ 50\,M$_\odot$.

\begin{figure}[!h]
\begin{center}
\psfig{figure=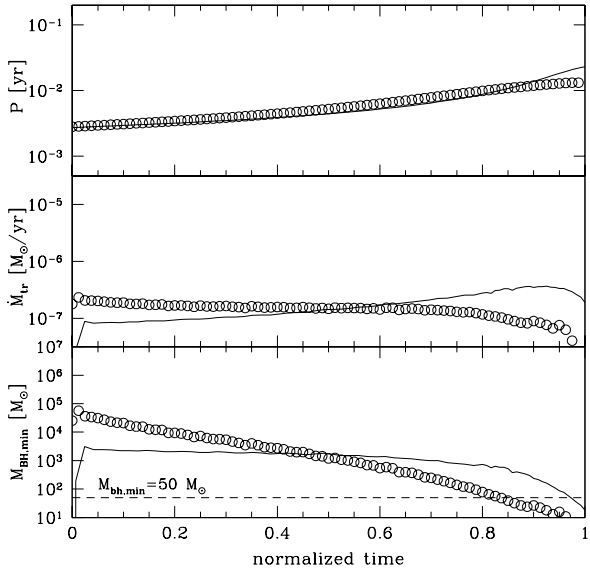,width=3.2in}
\caption{Comparison of orbital period (top), mass transfer rate (middle), and derived minimum
BH mass for transient behavior (bottom), for mass-transfer sequences with 10\,M$_\odot$ donors at ZAMS for a 1000\,M$_\odot$
BH (solid) and a 10\,M$_\odot$ BH (open circles with reduced time resolution for best clarity of the figure). It is evident that results are insensitive to the BH mass used in the mass-transfer calculations.}
\label{fig:mt}
\end{center}
\end{figure}

%\vspace{-0.5cm} 

Before drawing firm conclusions from our results, we examine whether the derived $M_{\rm BH,min}$ values are affected by the fact that we have used an IMBH for the calculations. We have repeated our sequences for a stellar-mass BH (10\,M$_\odot$) and found no qualitative or significant quantitative differences, as is evident in Figure 2 for two example sequences with a 10\,M$_\odot$ donor. These deviations occur because the mass ratio is close to unity in one case, and lead to {\em higher} values of $M_{\rm BH,min}$ for BH masses comparable to the donor masses.  

Given the qualitatively different results for donors more or less massive than $\simeq 5$\,M$_\odot$, we consider their relative probabilities as {\em Roche-lobe filling BH binary companions} in young star-forming regions where most ULXs are found. From basic results of stellar dynamics and preliminary calculations of our own (Ivanova, Kalogera, \& Belczynski, in preparation) we find that relatively massive companions in are favored for a number of reasons: (i) BHs sink by dynamical friction to the center of young star-forming regions, as do massive stars, and therefore there is more of them in the BH's vicinity; (ii) massive stars have a higher cross section for capture by a BH and, if exchange into binaries is relevant, lower-mass objects are generally ejected in the interaction; (iii) such stellar interactions strongly favor orbital periods in excess of $\sim 100$\,d, so even,  if a low-mass companion were present at some point in the BH dynamical lifetime, it would not fill its Roche lobe in young clusters (requires orbital periods shorter than $\simeq 1$\,d). 

On the other hand, such relatively massive donors to stellar-mass BH could drive thermal--timescale mass transfer and therefore produce {\em persistent} X--ray sources~\citep{King}. Thus we conclude that, if a substantial fraction of ULXs prove to be transient, then IMBH accretors could be favored.

\section{DISCUSSION} 

We have calculated mass transfer driven by relatively massive stars ($5-20$\,M$_\odot$) in BH binaries likely in young
stellar environments, and derived a minimum BH mass for transient behavior that in the majority of relevant cases is in excess of $50$\,M$_\odot$. This provides an observational diagnostic that could allow us to distinguish between stellar--mass ($\lesssim
20$\,M$_\odot$) and intermediate--mass BH binary models for ULXs.
We note that in old populations of ellipticals both classes of sources
are expected to be transient \citep{PB,K2002}. Hence transient
behavior cannot be used as an observational diagnostic in old stellar
systems (ages in excess of $10^{8}$\,yr).

So far there is only one candidate for a transient ULX. One
($L_X\simeq 1.1\times 10^{40}$\,erg\,s$^{-1}$) is in a starburst
galaxy NGC\,3628 \citep{S2001}. It may be associated with a
ROSAT X-ray source (so the position is not well constrained) that
faded below the sensitivity limit by a factor of more than 27 and
reappeared in {\em Chandra} observations. 

Current scenarios for IMBH involve formation in young stellar
clusters. One possibility invokes repeated black hole mergers~\citep{MH}, although
gravitational radiation recoil~\citep{RR} could prevent this by
ejecting merger products from the cluster. Another idea invokes
runaway collisions of massive stars and eventual collapse of the
massive remnant~\citep[provided that stellar winds do not decrease the
mass of the collision product; see][]{PZMcM}. Then an IMBH may form within the
lifetime of the most massive stars ($3$\,Myr), and it may acquire a binary companion within a cluster relaxation time after BH formation (at $\lesssim 10$\,Myr), when stars as massive as $\simeq 20$\,M$_\odot$ are still present. At that time binary separations would still be wide, favoring
the formation of IMBH binaries with orbital periods longer than about 100\,d. We are currently studying the dynamical evolution of an IMBH in young clusters and the characteristics of IMBH binaries and we expect to present our results in the near future (Ivanova, Kalogera, Belczynski 2003, in preparation).

It is important to realize that the transient behavior we discuss must occur on timescales far longer than an observer's lifetime. Therefore
tests for transient behavior must be carried out in a statistical sense. Any quantitative
statement ultimately depends on the duty cycle and the outburst
duration. It is also important to remember
that such a detection would imply a much larger number $\sim 1/d \ga
100 - 1000$ of quiescent systems of the same type.

\vspace{-0.5cm} 

\acknowledgments We thank J.~Miller, R.~Taam, and A.Zezas for useful
 discussions. VK and ARK acknowledge the hospitality and support of
 the Aspen Center for Physics (Summer 2002). This work is partially
 supported by a David and Lucile Packard Science \& Engineering
 Fellowship and a Chandra theory grant to VK, a NASA Summer Research
 fellowship fund to MH.  ARK gratefully acknowledges a Royal Society
 Wolfson Research Merit Award, and the hospitality and support of the
 Theoretical Astrophysics Group at Northwestern University.

\clearpage
%%%%%%%%%%%%%%%%%%%%%%%%%%%%%%%%%%%%%%%%%%%%%%%%%%%%%%%%%%%%%%%%%%%%%%%
%% Table 1. (table1)

%\clearpage

 %\centerline{\psfig{figure=f1.eps,angle=0,width=5.5in}} 
 %\figcaption{Average signal-to-noise degradation factor in pulsar search code versus survey integration time for PSR~B1913+16 and PSR~B1534+12.\label{fig:fvst}}

\end{document}